\documentclass[prb,preprint,showpacs,showkeys,amsmath,amssymb,superscriptaddress,10pt]{revtex4}

\usepackage{graphicx}
\usepackage{dcolumn}
\usepackage{graphicx}
\usepackage{epsfig}
\usepackage{mathtools}
\usepackage{amssymb}
\usepackage{natbib}
\usepackage{xcolor}
\usepackage{tikz}
\usepackage{calc}
\usepackage[caption=false]{subfig}
\usepackage{setspace}

\usepackage{frcursive}

\makeatletter
\newlength\@SizeOfCirc%
\newcommand{\CricArrowRight}[1]{%
    \setlength{\@SizeOfCirc}{\maxof{\widthof{#1}}{\heightof{#1}}}%
    \tikz [x=1.0ex,y=1.0ex,line width=.15ex, draw=black]%
        \draw [->,anchor=center]%
            node (0,0) {#1}%
            (0,1.2\@SizeOfCirc) arc (85:-240:1.2\@SizeOfCirc);%
}%
\makeatother

\begin{document}

\title{Magneto-Crystalline Anisotropy  of Fe, Co and Ni slabs from Density Functional Theory and Tight-Binding models.}

\author{L. Le Laurent}
\affiliation{SPEC, CEA, Universit\'e Paris-Saclay, CEA Saclay 91191 Gif-sur-Yvette, France}

\author{C. Barreteau}
\affiliation{SPEC, CEA, CNRS, Universit\'e Paris-Saclay, CEA Saclay 91191 Gif-sur-Yvette, France}

\author{T. Markussen}
\affiliation{Synopsys Denmark, Fruebjergvej 3, Postbox 4, DK-2100 Copenhagen, Denmark}

\date{\today}
\begin{abstract}
We report magneto-crystalline anisotropy (MCA) calculations of Fe, Co and Ni slabs of various thicknesses and crystallographic orientations from two Density Functional Theory codes based either on a plane wave or a local atomic basis set expansion and a magnetic tight-binding method. 
We analyze the evolution of the MCA with the number of layers of the slabs. The decomposition of MCA into contributions of atomic sites helps understanding the oscillatory behaviour of the MCA with the slab thickness and highlights the role of finite size effects. We also identify some specific systems with enhanced MCA.
A k-space as well as a band-filling analysis show very rich features of the MCA that could be used to tailor systems with enhanced magnetic properties. Finally this work can serve as a benchmark for MCA calculations.
\end{abstract}

\maketitle

\section{Introduction}

Spin-orbit coupling driven phenomena are subject of intense studies since it is at the origin of many fundamental physical effects. Magneto-crystalline  anisotropy (MCA) is one of the important consequence of the coupling between the spin and the orbital moment of the electron\cite{Bruno1989}. It is a property of central interest for both fundamental and practical reasons. 
MCA is characterized by the dependence of the energy of a magnetic system on the orientation of the magnetization with respect to the crystallographic structure of the material. The axis (or plane) corresponding to the minimum of energy is the so-called easy-axis (plane). The magnetic energy landscape has many physical implications for example on the thermal stability of magnetic nanoparticles, and for technological application the development of materials with large uni-axial anisotropy is often requested\cite{Dieny2017}. In addition tunability is often a desired functionality and therefore it is essential to analyze and understand the main parameters governing the magnetic anisotropy in order to find ways to control MCA in an efficient way. 

From a computational point of view the calculation of the MCA is \textit{a priori} straightforward since one only needs to compute the total energy for  different magnetization orientations. However due to the smallness of energy differences and other technical details, the determination of the MCA is numerically delicate. In addition, besides the total MCA it is also essential to have efficient numerical ways to decompose the MCA as a sum of local contributions in heterogeneous systems presenting surfaces, interfaces or any type of defects. Several options have been proposed either based on the (second order) quantum mechanical perturbation (2PT) theory\cite{Bruno1989,Wang1993,Autes2006} or on the force theorem (FT) \cite{Weinert1985,Daalderop1990,Wang1996}. Both approaches allow a local site (and orbital) analysis of the MAE\cite{Yang2016,Li2013}. However, the domain of validity of the FT is \textit{a priori} larger than 2PT\cite{Blanco-Rey2019}. 

In this paper we present a series of Density Functional Theory (DFT) and Tight-Binding (TB) calculations based on the FT to evaluate the MCA of Fe, Co and Ni slabs of various thicknesses and crystallographic orientations.
We used two very different DFT codes: Quantum espresso (QE \cite{Giannozzi2009,Giannozzi2017}) based on plane wave expansion of the wave functions, QuantumATK (QATK\cite{QATK2019,Smidstrup2017}) based on a linear combination of atomic-like orbitals. We also compare our results with a semi-empirical magnetic TB method \cite{Barreteau2016}. In order to understand the agreements or discrepancies between the different methods and extract general trends we provide a local site analysis which allows to extract the surface contribution to the MCA. In addition since the MCA is extremely sensitive to tiny details of the band structures we have performed $k$-space as well band-filling analysis providing important information that can be used to tune the MCA.

\section{Methodology}

In the following section we will present the main ingredients and technical details of the methods used to determine the MCA.

\subsection{Density Functional Theory (DFT)}

We have performed DFT calculations using the generalized gradient approximation (GGA) in the same Perdew, Berke, Ernzherof (PBE \cite{Perdew1996})parametrization but based on radically different expansions of the valence electrons wave-functions: QE\cite{Giannozzi2009,Giannozzi2017} uses plane waves while QATK\cite{QATK2019} uses localized atomic-like orbitals. 
With QE we used ultrasoft pseudotentials\cite{Vanderbilt1990} and the size of the basis, controlled by the energy cutoff, was taken equal to 30Ry and 300Ry for the wave function and the charge density respectively, while with QATK we used the norm-conserving PseudoDojo pseudopotentials \cite{VanSetten2018} and a High basis set\cite{QATK2019}.
As explained in greater details in Sec.\ref{sec:FT} we have applied the Force Theorem (FT) to evaluate the MCA and its local components. This approach is based on a three steps process\cite{Li2014,Smidstrup2019}: i) a self-consistent (scf) calculation with a scalar relativistic pseudopotential (without spin-orbit coupling), followed by ii) a non-self-consistent (nscf) calculation with a fully relativistic pseudopotential  including spin-orbit coupling starting from the scf electron density rotated to the specified magnetization orientations. Finally iii) the MCA is obtained from the variation of the band energy term. In most of the calculations (unless explicitly stated) the scf loop is performed with a $25\times 25$ $k$-point sampling of the two dimensional Brillouin zone while the nscf calculations are performed with a denser sampling of $50\times 50$ $k$-points in the full Brillouin zone. Marzari-Vanderbilt cold smearing with a broadening of 0.05eV has been used. 
The local quantities are obtained by slightly different procedures: 
QE projects the wave-functions onto orthogonalized atomic pseudo-wave-functions in a Lowdin manner while QATK naturally projects onto the atomic-like orbitals (used as a basis) in a Mulliken manner\cite{QATK2019}.

\subsection{Magnetic Tight-binding (TB)}

We have also used an $spd$ semi-empirical tight-binding method\cite{Barreteau2016} where the spin magnetism is taken into account via a Stoner-like potential $V^{\text{Stoner}} = -1/2 I^{\text{Stoner}}  \mathbf m.\mathbf \sigma$ and the spin-orbit coupling potential acting on $d$ orbital is written $V^{\text{SOC}} = \xi  \mathbf L.\mathbf S$.
The TB parameters of the model are determined by fitting to DFT results (see Tab. \ref{tab:TB-param} for the SOC and Stoner parameters).
The control parameters: number of $k$-points, broadening, convergence threshold are the same as for the DFT calculations. 

\begin{center}
\begin{table}[ht]
\begin{tabular}{|c|c|c|c|}
\hline
 & Fe  & Co  & Ni  \\
\hline
$I^{\text{Stoner}}$ (eV) & 0.95 & 1.10 & 1.05 \\
\hline
$\xi^{\text{SOC}}$ (eV) & 0.06 & 0.08 & 0.10 \\
\hline
\end{tabular}
\caption{\label{tab:TB-param} Fe,Co, Ni Stoner and spin orbit coupling parameters used in the magnetic TB model.}
\end{table}
\end{center}

\subsection{Force Theorem}
\label{sec:FT}

The MCA was calculated by making use of the Force Theorem (FT) which is valid in the case of "not too large" spin orbit coupling  (see next section for the validation). Within the FT procedure the MCA is obtained as the difference of band energy after a single diagonalization. In a periodic system the eigenvalues (and eigenfunctions) are labelled by a $k$ vector and a discrete number $n$ of bands such that the MCA can be written:

\begin{equation}
\text{MCA}^{\text{FT}} = \sum_{k} \sum_n f_{k,n}^1 \varepsilon_{k,n} (\hat{\mathbf m}_1)-\sum_{k } \sum_n f_{k,n}^2 \varepsilon_{k,n} (\hat{\mathbf m}_2)
\end{equation}

where $\varepsilon_{k,n} (\hat{\mathbf m})$ are the eigenvalues obtained after a single diagonalization of the Hamiltonian including SOC but starting from well-converged charge/spin density of a self-consistent calculation without SOC rotated in a given spin orientation $ \hat{\mathbf m}$. $f_{k,n}=f(\varepsilon_{k,n}-E_{\text{F}})$ are the filling factors.
Note that within this approach the Fermi level is different for the two orientations. However to decompose the total MCA onto local components $\text{MCA}_i$  it is necessary to adopt the grand canonical description\cite{Bonski2010,Li2013}:

\begin{equation}
\label{eq:MCAi}
\text{MCA}^{\text{FT}_{\text gc}} = \int f(E) (E-E_{\text{F}})\Delta n(E)\, \mathrm{d}E= 
\sum_i \int f(E) (E-E_{\text{F}})\Delta n_i(E)\, \mathrm{d}E  = \sum_i \text{MCA}_i
\end{equation}

where $n_i(E)$ is the density of states projected on site $i$ and $\Delta n_i(E)$ the difference between the two spin orientation $\hat{\mathbf m}_1$ and $\hat{\mathbf m}_2$.
Here $E_{\text{F}}$ is the Fermi level of the system without SOC corresponding to a neutral system. However from a computational point of view it is also interesting to explore the behaviour of the MCA when varying the Fermi level away from the neutrality point\cite{Blanco-Rey2019} as will be illustrated further below.

\section{MCA of Co, Fe and Ni slabs}

We have considered the three $3d$ transition metal ferromagnetic elements Fe, Co and Ni in their equilibrium structure: body centered cubic (bcc) for Fe, hexagonal close pack (hcp) for Co and face centered cubic (fcc) for Ni. In the case of Co we have also considered its fcc structure since this element can easily adopt this structure when grown on a substrate. 
For the sake of comparison between the codes we have ignored surface relaxation and therefore the calculations have been performed on exactly the same structures and with the same computational parameters. 
The structural parameters are summarized in Tab. \ref{tab:lattice}. 

\begin{center}
\begin{table}[ht]
\begin{tabular}{|c|c|c|c|c|}
\hline
 & Fe bcc  & Co fcc  & Co hcp  & Ni fcc  \\
\hline
Lattice parameter (\AA) & a = 2.8665 & a = 3.5447 & a = 2.5071 & a = 3.5249 \\
   & & & c = 4.0686 & \\
\hline
\end{tabular}
\caption{\label{tab:lattice} Fe,Co, Ni lattice parameters used in this work.}
\end{table}
\end{center}

\begin{figure}[!ht]
\includegraphics[width=8cm]{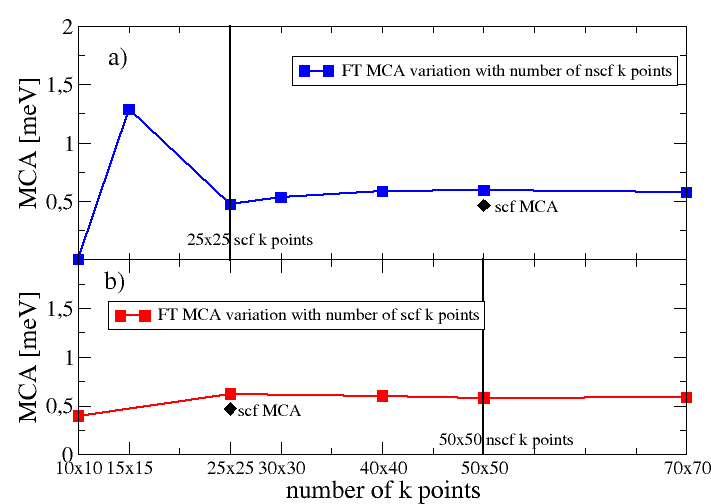}
	\caption{ \label{fig:MCA-vs-nk} Results of TB MCA calculations: a) Total MCA (using FT) of a Co  hcp $(0001)$  15 layer slab with respect to the number of nscf $k$ points (including SOC), the number of scf $k$ points (collinear spin without SOC) being fixed at $25 \times 25$ (vertical straight line) . b) Total MCA (using FT) of a  Co  hcp  $(0001)$  15 layer slab with respect to the number of scf $k$ points, the number of nscf $k$ points being fixed at $50 \times 50$ (vertical straight line). The $\blacklozenge$ indicates the results of a full scf calculation, {\sl i.e.} not making use of the FT. }
\end{figure}

In all our calculations  the MCA is defined as  $\text{MCA} = E_{\parallel} - E_{\perp}$ (per surface unit cell)  where $\parallel=\hat{\mathbf m}_1=x$ is in the plane of the slab and $\perp=\hat{\mathbf m}_2=z$  is perpendicular to the surface of the slabs. We checked that the in-plane anisotropy is extremely small and therefore the choice of the $x$ axis in the surface plane does not influence the quantitative value of the MCA. With this convention a positive MCA means an out-of-plane easy axis.  We have also checked (within the FT approach) the convergence with the number of $k$ points of the scf (no SOC) and nscf (with SOC) calculations. In Fig. \ref{fig:MCA-vs-nk} is shown the evolution of the total MCA for a  Co hcp$(0001)$ 15 layer slab  while varying the number of (scf or nscf) $k$ points. It comes out that the MCA is indeed more sensitive to the number of nscf $k$ points than to the number of scf ones and typically $25 \times 25$ and $50 \times 50$ scf and nscf $k$ points respectively are sufficient to achieve converged MCA. Finally, to validate the FT approach we have also calculated the MCA from a scf calculation including SOC with $50 \times 50$ kpoints. The agreement with FT is excellent (see $\blacklozenge$ in Fig.\ref{fig:MCA-vs-nk}), the error is estimated to be of the order of $10^{-4}$eV=$0.1$meV for a system containg 15 atoms in the unit-cell. 

\subsection{Evolution of the MCA with the slab thickness}

\begin{figure}[!ht]
\subfloat[Co TB ]{\includegraphics[width=6cm]{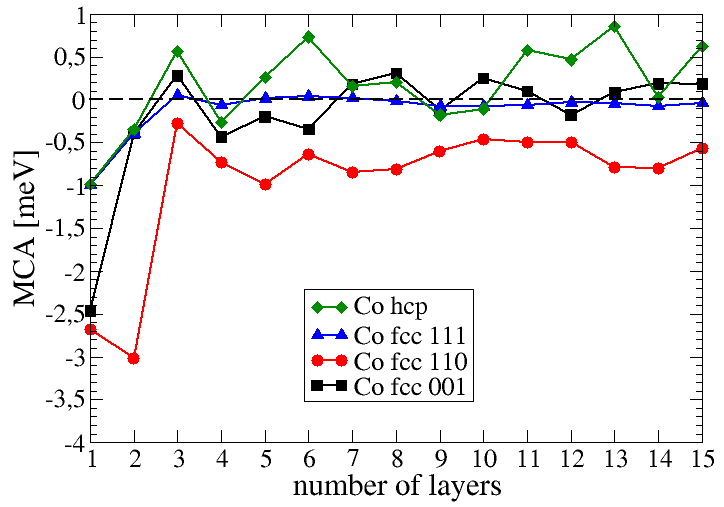}}
\subfloat[Co QE ]{\includegraphics[width=6cm]{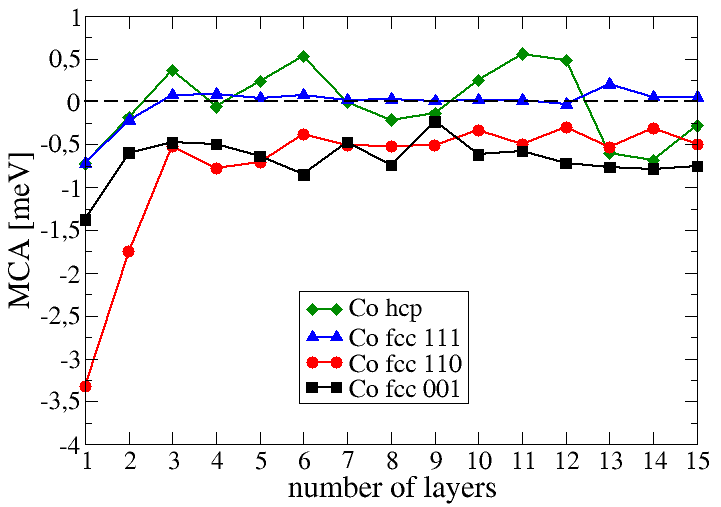}}
\subfloat[ Co QATK]{\includegraphics[width=6cm]{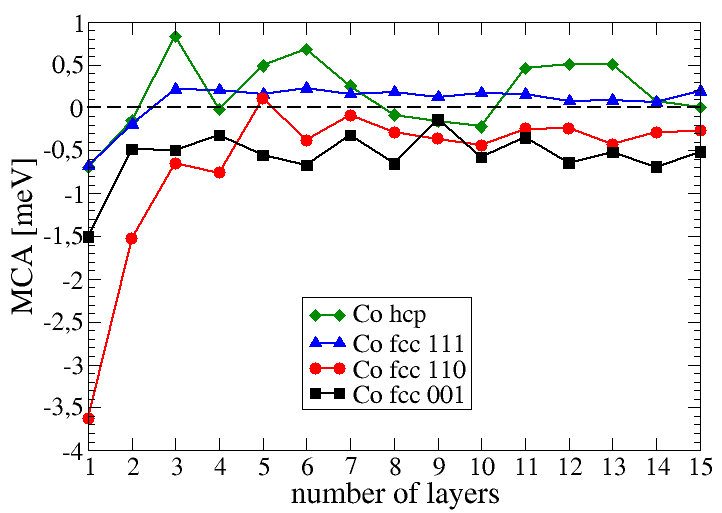}}

\subfloat[ Fe TB]{\includegraphics[width=6cm]{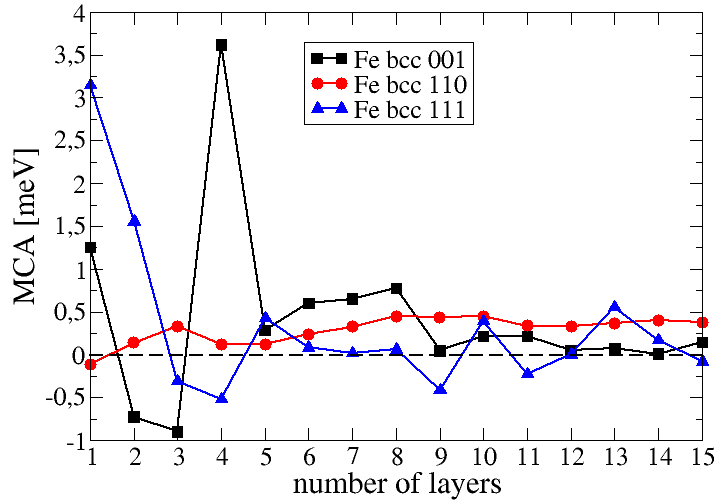}}
\subfloat[Fe QE ]{\includegraphics[width=6cm]{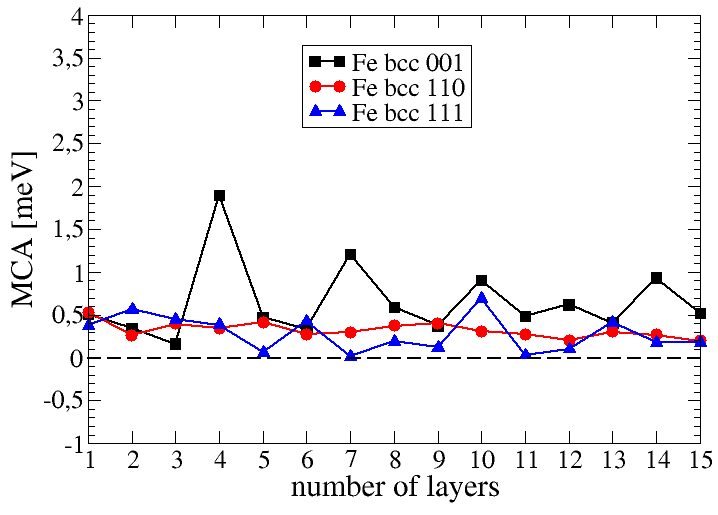}}
\subfloat[Fe QATK ]{\includegraphics[width=6cm]{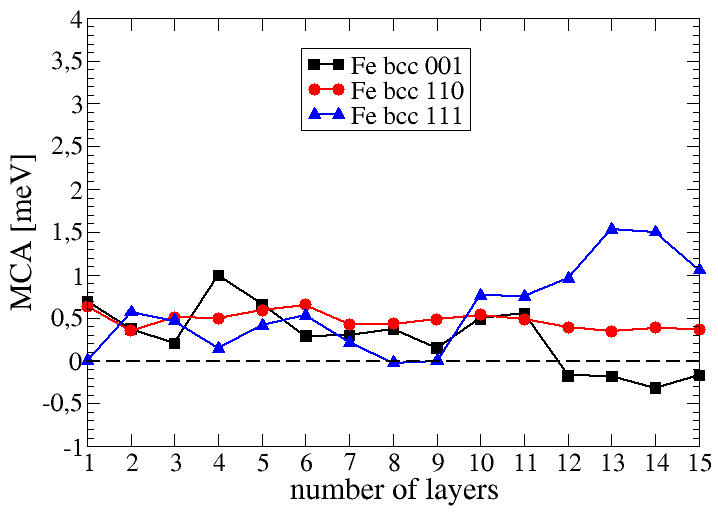}}

\subfloat[ Ni TB]{\includegraphics[width=6cm]{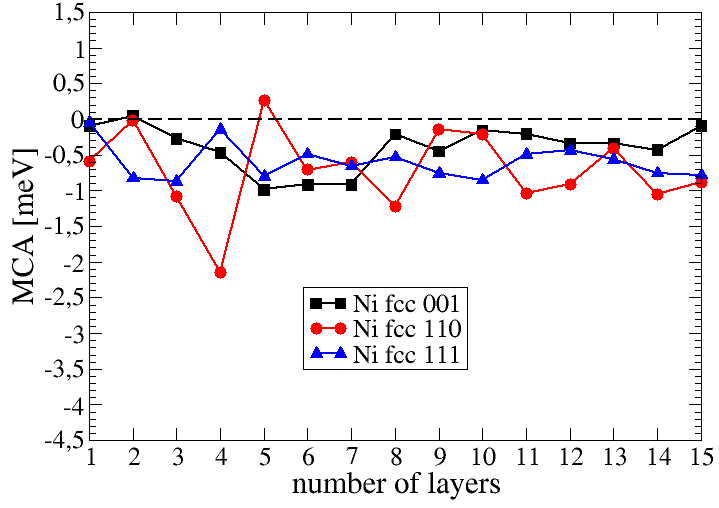}}
\subfloat[ Ni QE]{\includegraphics[width=6cm]{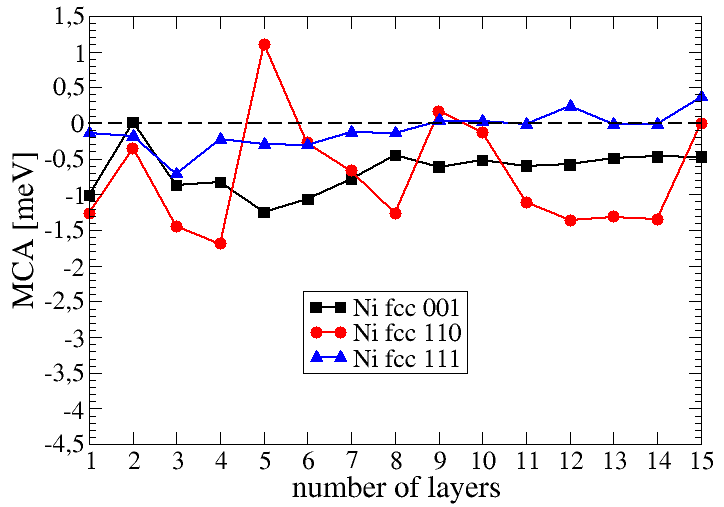}}
\subfloat[Ni QATK ]{\includegraphics[width=6cm]{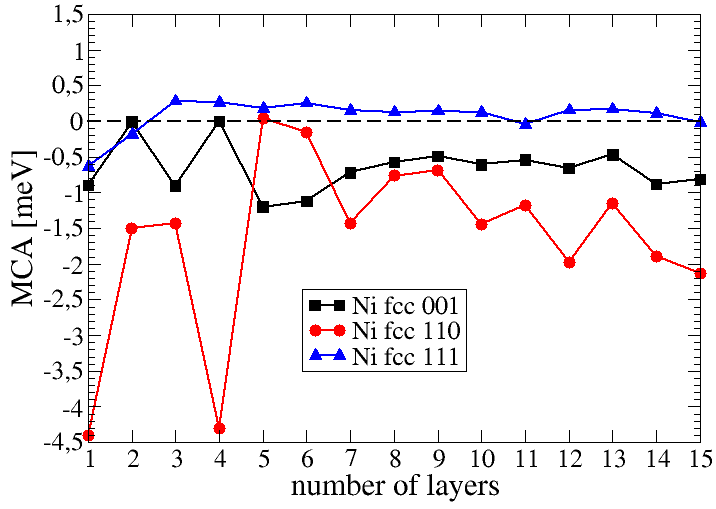}}
	\caption{\label{fig:MCA-N} Total $\text{MCA}$  as a function of the number of layers of the slabs  for Co (TB (a), QE (b), QATK (c)), Fe (TB (d), QE (e), QATK (f)) and Ni (TB (g), QE (h), QATK (i)).}
\end{figure}

We have performed total MCA calculations of slabs for the three main cubic $(001)$, $(110)$ and $(111)$ crystallographic orientations and $(0001)$ for hcp. The slab thicknesses have been varied from one to fifteen layers. The results are shown in Fig. \ref{fig:MCA-N}. The overall qualitative agreement between the different codes is satisfactory and the general trends are reasonably well reproduced, in particular the sign of the MCA is generally identical (positive for Fe, negative for Ni and close to zero for Co). As expected, for very small thicknesses large anisotropies as well as large amplitudes of oscillations are observed for $1\le N\le5$. In addition, for specific thicknesses and slab orientations some accidents with maxima or minima of MCA are observed. This is the case of the three-layer slab of Co hcp$(0001)$ (in agreement with Ref. \onlinecite{Yang2016}), four-layer slab of Fe bcc$(001)$ (already observed in Ref.  \onlinecite{Li2013}) or the four and five layer slabs of Ni$(110)$.

For thicknesses above $N=5$ the MCA generally stabilizes but in several cases an erratic oscillating behaviour is observed up to 15 layers. This behaviour is usually attributed to the so-called quantum-well states\cite{cinal2003,Gimbert2012}. However, as already described in our previous publication\cite{Li2013}, it is only for very large thicknesses (above 30 layers) that the effect of quantum states oscillations can clearly be identified. This will be illustrated in Sec. ~\ref{sec:thickCo} for Co hcp$(0001)$ slabs.  In fact we prefer to speak about finite size rather than true quantum well states effects.

\subsection{Site-resolved MCA}

\begin{figure}[!ht]
\subfloat[Co TB ]{\includegraphics[width=6cm]{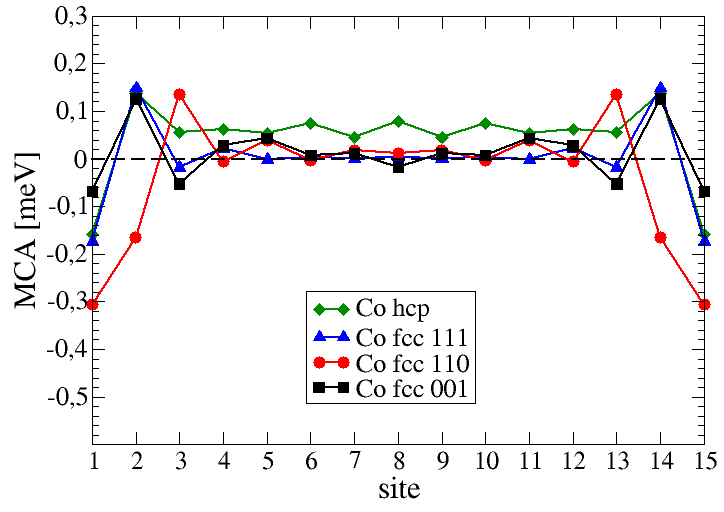}}
\subfloat[ Co QE ]{\includegraphics[width=6cm]{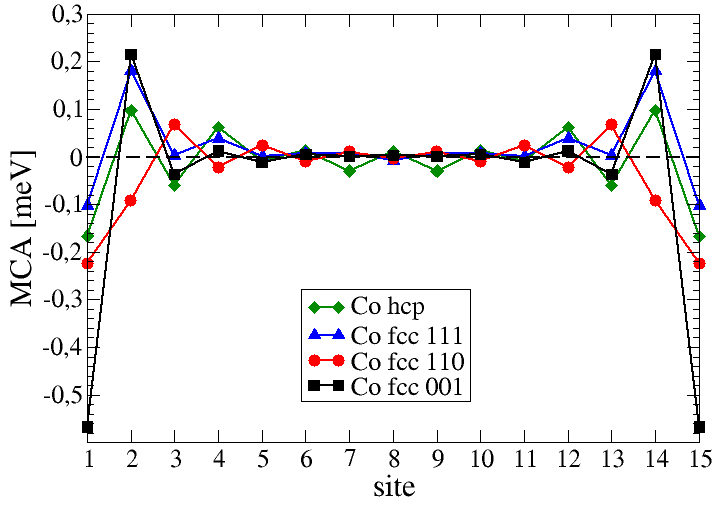}}
\subfloat[ Co QATK]{\includegraphics[width=6cm]{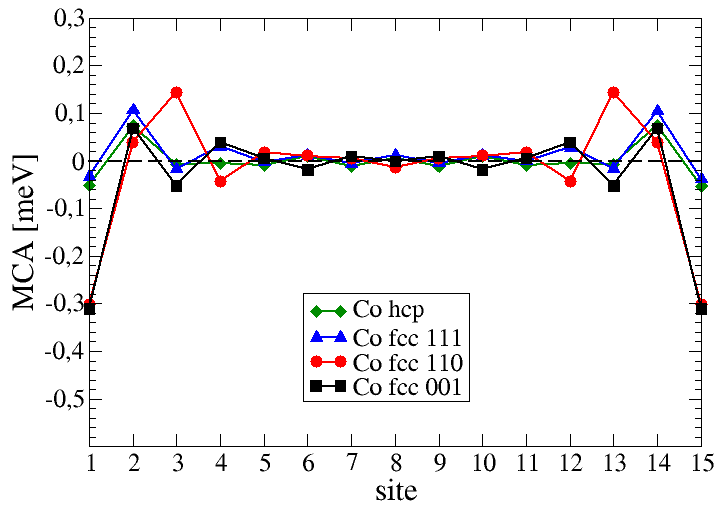}}

\subfloat[ Fe TB]{\includegraphics[width=6cm]{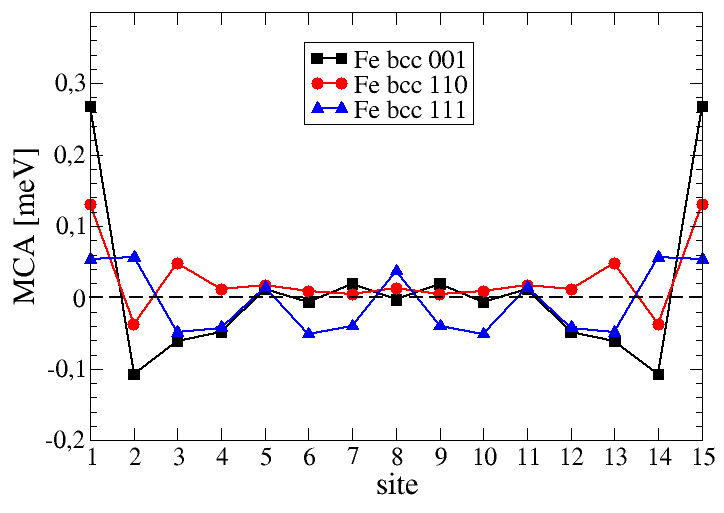}}
\subfloat[ Fe QE]{\includegraphics[width=6cm]{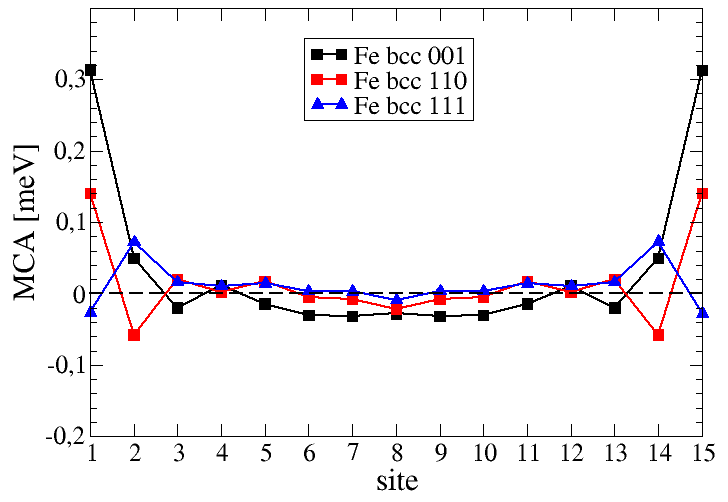}}
\subfloat[ Fe QATK]{\includegraphics[width=6cm]{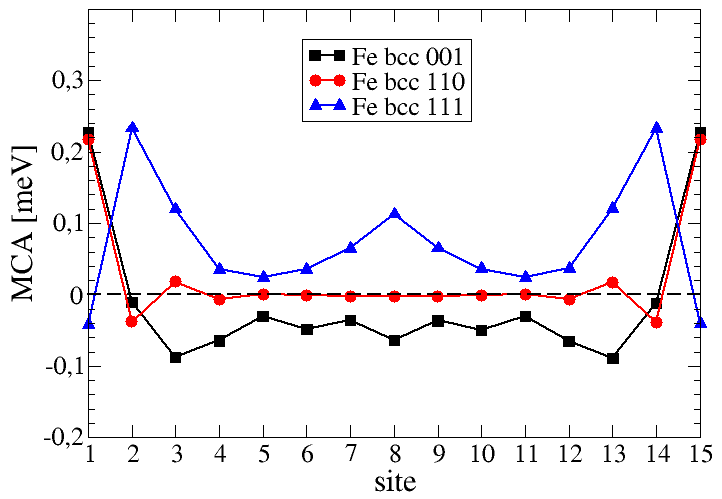}}

\subfloat[Ni TB ]{\includegraphics[width=6cm]{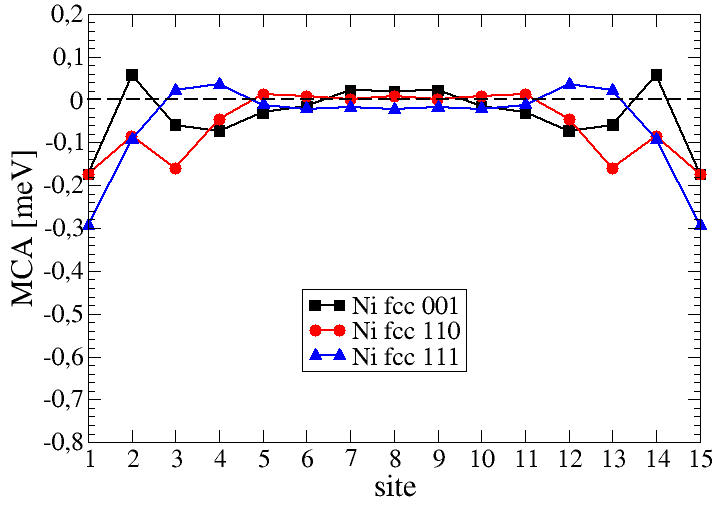}}
\subfloat[Ni QE ]{\includegraphics[width=6cm]{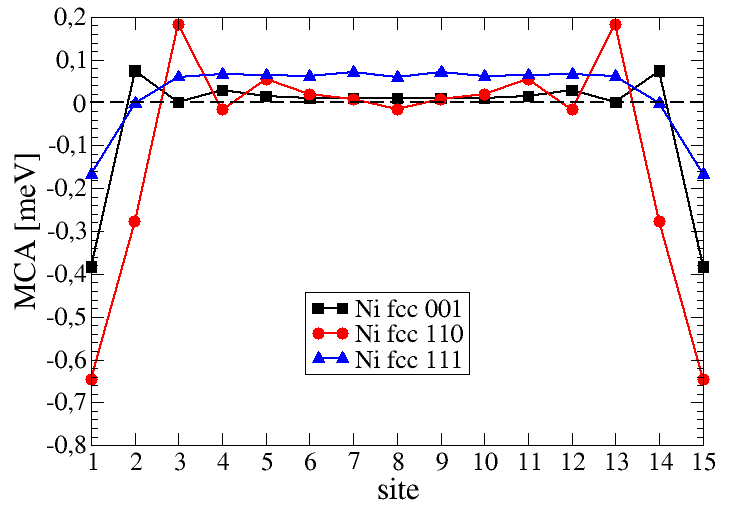}}
\subfloat[Ni QATK ]{\includegraphics[width=6cm]{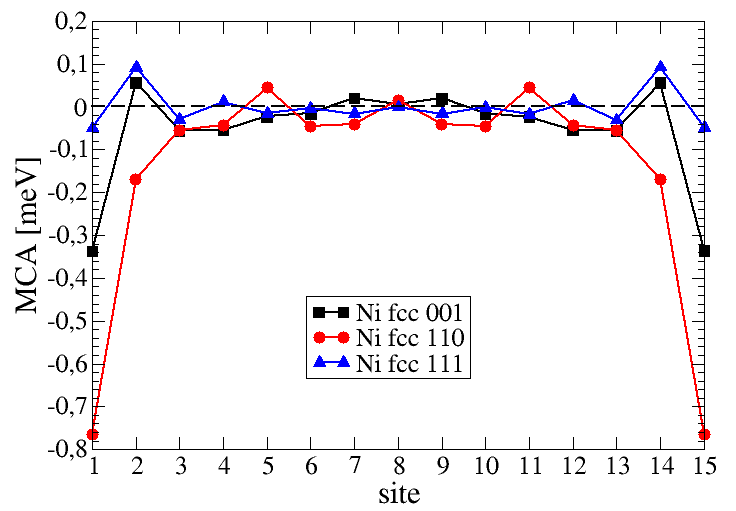}}
	\caption{\label{fig:MCA-site} Layer resolved MCA (see Eq. \ref{eq:MCAi} )  for 15-layer slabs of  Co ( TB (a), QE (b), QATK (c)), Fe ( TB (d), QE (e), QATK (f)) and Ni ( TB (g), QE (h), QATK (i)).}
\end{figure}

The evolution of the total MCA  with the number of atomic layers can only be understood thoroughly via a layer-resolved MCA analysis. Indeed, when increasing the slab thickness the number of atoms with a bulk-like environment is increasing while the number of surface-like atom is constant. Therefore the total MCA($N$) can be decomposed into a bulk contribution that should increase linearly (if the bulk MCA is non-zero) with the number of layers and a constant surface term. However, within this simplistic picture, finite size and quantum well states effect are neglected. In Fig. ~\ref{fig:MCA-site}, we have decomposed the MCA of all the 15-layer slabs considered. From these results a clear distinction can be made between the two (or three) outermost layers and the central part of the slab. Clearly the MCA surface component is negative (in-plane) for Cobalt and Nickel and positive (out-of-plane) for Iron. Note also that the sub-layer often counterbalances the outermost layer. In the case of Cobalt (hcp$(0001)$, fcc$(111)$ and fcc$(001$) the outermost and sub-layer MCA sum up to almost zero. 

In the central part, the MCA generally converges with small oscillations towards the bulk value. Save for hcp all the crystallographic structures considered are  cubic, hence the bulk MCA is
expected to be very small, more precisely experimentally it is found  of the order of $\mu$eV for Fe(bcc) and Ni (fcc) and around 60$\mu$eV in Co (hcp) \cite{Kubler_book}.
In consequence only the surface contribution (in fact twice the surface contribution since each slab contains two surfaces) should remain in $\text{MCA}(N)$ for large enough slab thicknesses $N$. This overall general picture is well obeyed by the  MCA($N$) curves of Fig.\ref{fig:MCA-N} but there are still several systems where the stabilization of MCA($N$) is very slow and large deviations are observed. This departure from the general simple picture is due to two main reasons: the strength of the perturbation induced by the presence of surfaces and finite size effects. The strength of the surface perturbation is essentially related to the number of neighbours that are lost at the surface. The more open the surface the strongest the perturbation. This rule of thumb is well obeyed and the MCA of fcc$(111)$ or bcc$(110)$ slabs converges fast towards the bulk (zero) value (see Fig.\ref{fig:MCA-site}) which can also be seen from the fast stabilization of MCA($N$) (see Fig.\ref{fig:MCA-N}). In contrast for bcc$(111)$ and fcc$(110)$ the MCA is perturbed over at least four layers. The QW states in metallic ultra-thin films modulate the density of states at the Fermi level and create periodic oscillations with the film thickness that are related to the bulk Fermi wave number in the $z$ direction perpendicular to the film. However this {\sl a-priori} simple rule is obscured by the complexity of the band structure in transition metals for which the Fermi surface can be very intricate.  In practice it is very difficult to predict their quantitative influence\cite{cinal2003,Gimbert2012}.

\subsection{Co hcp $k$-resolved MCA and band-filling analysis}

\begin{figure}[!ht]
\includegraphics[width=12cm]{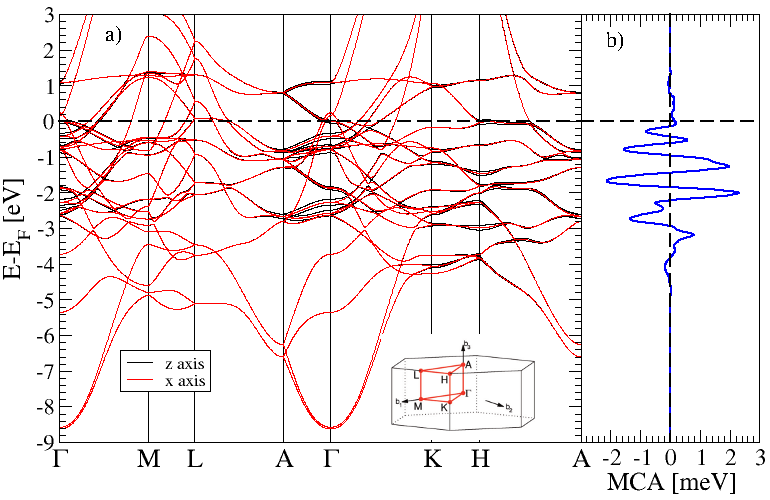}
	\caption{\label{fig:bandCohcp} a) TB+SOC band structure of bulk Co hcp along $\Gamma$-L-A-$\Gamma$-K-H-A path (see  Brillouin zone in inset). In black(red) is shown the band structure corresponding to an out-of-plane (in-plane) magnetization . b) Total MCA as a function of the position of  the Fermi leve.}
\end{figure}

To get further insight it is also interesting to analyze the MCA in the $k$-space since this allows to identify the regions contributing to the MCA. 
Let us focus on the bulk Co hcp. In Fig. \ref{fig:bandCohcp}a) we have plotted the TB band structure of Co hcp along a high symmetry path for two directions of the spin: $z$ ($c$ axis) and $x$ ($a$ axis). Due to the smallness of SOC, the two band structures are almost identical apart from specific regions with relatively flat bands that are split when the magnetization is along $z$ and not split when it is along $x$. These regions should contribute the most to the MCA for band filling such that these bands would cross the Fermi level. To illustrate this idea we have calculated the total MCA as a function of the Fermi level which corresponds to changing the band-filling. The curve plotted in Fig.\ref{fig:bandCohcp}b) is very instructive since it shows rapid variations of the MCA at energies corresponding to the position of these flat bands. Interestingly, small variations of the band filling can lead to drastic changes of the MCA of several meV with change of sign and therefore  of easy axis\cite{Blanco-Rey2019}.

Note that the total MCA is integrated over the whole Brillouin zone and the result is the sum of contributions coming from various regions favoring either in-plane or out-of-plane anisotropy. 
It is also worth mentioning that in TB the bulk MCA is slightly positive (0.05meV), while it is almost zero with QE and QATK.  In the present case TB is in better agreement with the experimental MCA\cite{Kubler_book} ($0.065$eV) than QE and QATK  but this might be fortuitous. This is essentially due to the position of bands  forming an inverted parabola around the $\Gamma$ point. These bands cross the Fermi level in TB while they are below the Fermi level in QE and QATK. One can also check  from Fig.\ref{fig:bandCohcp}b) that the MCA falls at almost zero 0.1eV above the Fermi level where the inverted parabola is filled.

\begin{figure}[!ht]
\subfloat[ ]{\includegraphics[width=9cm]{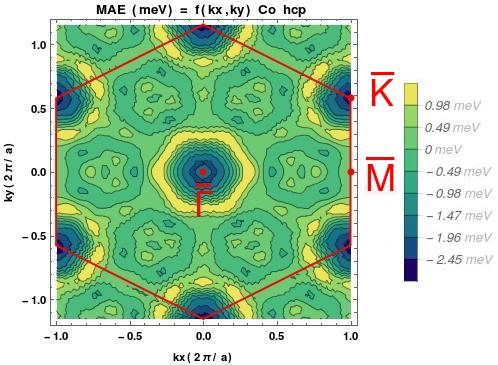}}
	\caption{\label{fig:maekxky} $(k_x,k_y)$ resolved MCA of bulk Co hcp (integrated along $k_{z}$). We can easily see that the most important contributions originate  from regions around the high symmetry points  $\bar{\Gamma}$ and $\bar{K}$.}
\end{figure}

We have then evaluated the MCA from a dense $100\times100\times100$ $k$-mesh TB calculation. Since we have in mind  the MCA of Co hcp$(0001)$ slabs we have integrated the MCA along the $k_z$ direction. The resulting MCA($k_{x}$,$k_{y}$) is shown in Fig. \ref{fig:maekxky} clearly evidencing the role of symmetry points around which the major contribution to the MCA is coming. $\bar{\Gamma}$ and $\bar{K}$ points (we adopt the 'bar' notation to indicate that it is the result of a projection along $k_z$) regions show a negative dip surrounded by a positive ring that leads to the overall positive total MCA. One should however note that the MCA around the $\bar{K}$-point originates essentially from the projection of a zone near the $H$-point. The rest of the $k$ space is almost flat with zero anisotropy. Note also that the $\bar{M}$ point does not play any special role. 

\subsection{Large thickness limit}
\label{sec:thickCo}

\begin{figure}[!ht]
\includegraphics[width=9cm]{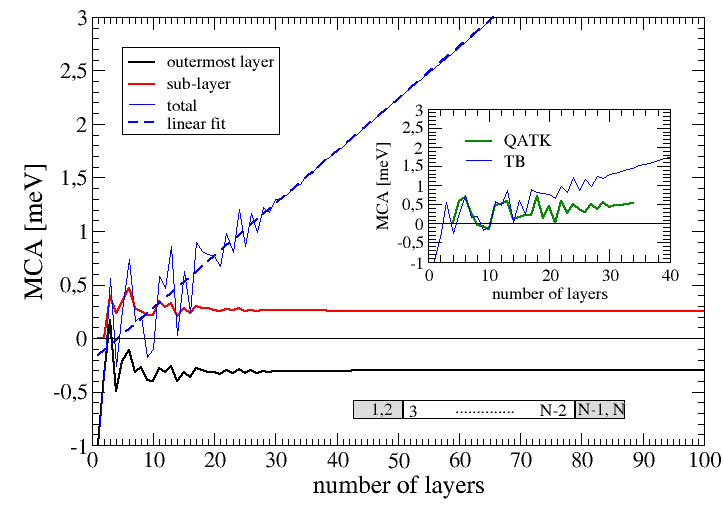}
	\caption{\label{fig:MCA100L} Total MCA (in blue) versus the number of layers of Co hcp$(0001)$ slabs obtained from the TB method. The outermost and sub-layer contributions are shown in black and red respectively. The outermost and sub-layer MCA includes two equivalent sites $(1,N)$ and $(2,N-1)$ respectively. The linear fit of the bulk MCA is shown in dashed blue. In the inset we have compared the TB results (in blue) to the QATK ones (in green) up to 35 layers. The lower slope is corresponding to the smaller bulk MCA value obtained with QATK.
}
\end{figure}

For $N$-layer slabs one expects the total MCA to be written as a constant surface term plus a linearly increasing bulk term: $\text{MCA}_{\text{tot}}(N)=2\times \text{MCA}_{\text{surf}}+N \times \text{MCA}_{\text{bulk}}$. Since Co hcp is the only system for which the bulk contribution is non zero (the other systems are cubic with extremely small MCA) we have calculated the total MCA for a series of Co hcp$(0001)$ slabs up to one hundred layers to check the general expected trend. The result is represented in Fig. \ref{fig:MCA100L} in which we have plotted the total MCA as a function of the number of layers. The linear behaviour is clearly demonstrated. However, until 30 layers strong oscillations are present and below 15 layers it is impossible to identify the linear scaling.  We have also extracted the contribution from the outermost layer (sites 1 and $N$) and sub-layer (sites  2 and $N-1$). Interestingly, although of opposite signs, the MCA of the outermost and sub-layer follow the same trend and the amplitude of their oscillations become almost negligible above 15 layers. It is also clear that the oscillations of the total MCA are dominated by the bulk between 15 and 30 layers. We have then fitted  $\text{MCA}_{\text{tot}}(N)$ by a linear formula $a+b N$ where $b$ gives the bulk contribution. We found $a=-0.21$meV and $b=0.049$meV which gives MCA$_{\text{bulk}}=0.049$meV and $\text{MCA}_{\text{surf}}=-0.105$meV. This is in perfect agreement with our previous estimation of the bulk MCA obtained from a purely bulk calculation (far less time consuming!). Note that the extraction of the surface contribution obtained from inverting the $\text{MCA}_{\text{tot}}(N)$ formula: $\text{MCA}_{\text{surf}}=\frac{1}{2}\big( \text{MCA}_{\text{tot}}(N)-N \times \text{MCA}_{\text{bulk}}\big)$   as one usually proceed for calculating surface energies  is not the best strategy because of the oscillating behaviour of the total MCA. A better solution is to sum the local contributions over the first outermost layers  obtained from a single slab calculation. Indeed the surface MCA can be more safely obtained from the following formula:

\begin{equation}
\label{eq:MCA-surf}
\text{MCA}_{\text{surf}}=\sum_{s=1}^{N_{\text{surf}}} \text{MCA}_s -N_{\text{surf}} \text{MCA}_{\text{bulk}}
\end{equation}

where $N_{\text{surf}}$ is the number of layers upon which the MCA is significantly "perturbed" ( (1, 2) in the case of Co hcp$(0001)$) and $ \text{MCA}_s$ is the contribution from layer $s$. Applying Eq. \ref{eq:MCA-surf} to Co hcp$(0001)$ one gets the same value -0.1meV as obtained from the linear fit procedure.

\subsection{Surface band-filling analysis}

\begin{figure}[!ht]
\subfloat[Co hcp$(0001)$ ]{\includegraphics[width=6cm]{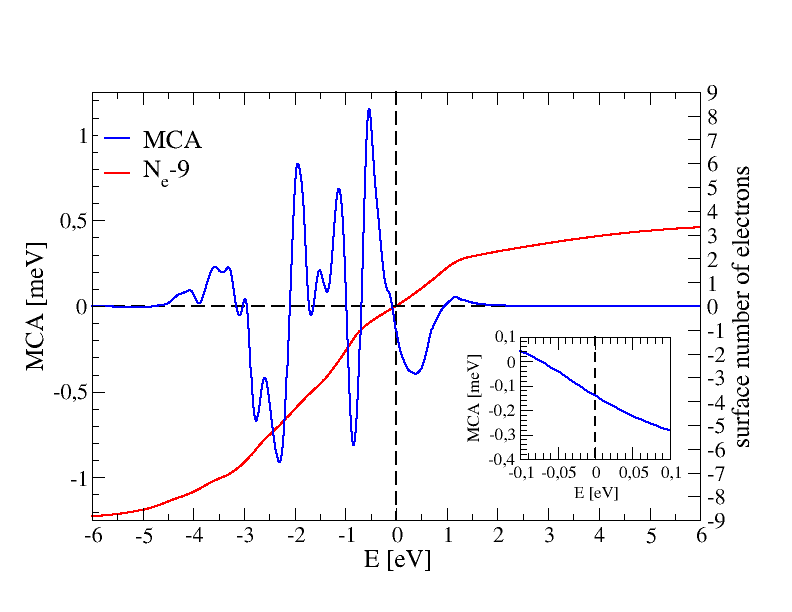}}
\subfloat[Fe bcc$(001)$ ]{\includegraphics[width=6cm]{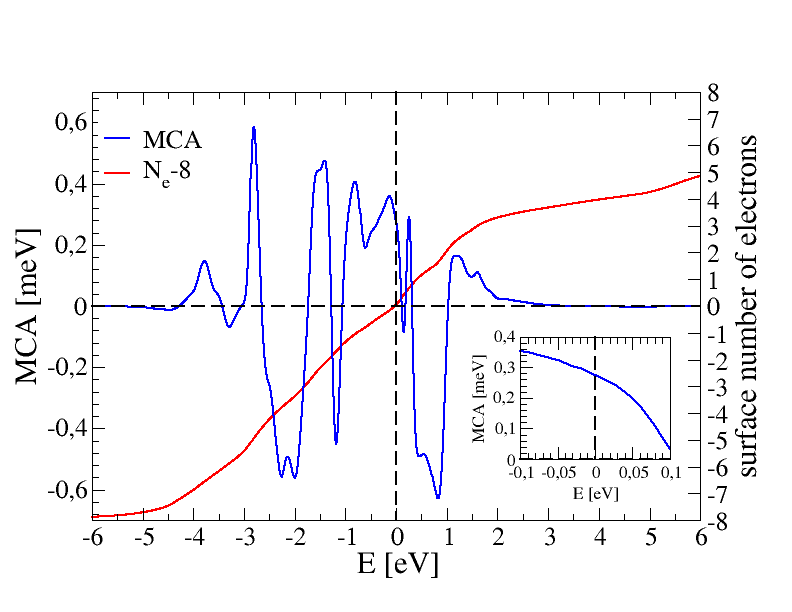}}
\subfloat[Ni fcc$(001)$ ]{\includegraphics[width=6cm]{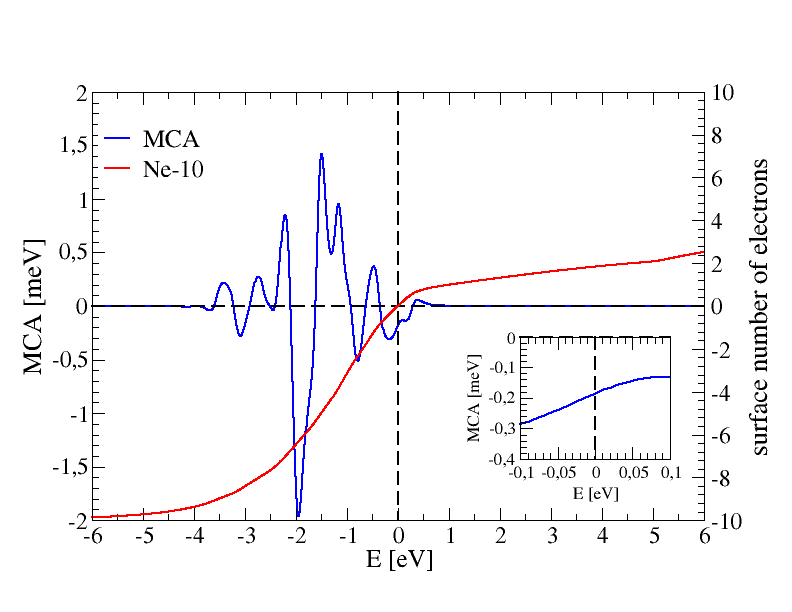}}

	\caption{\label{fig:MCA-filling}  MCA of the outermost layer of a 15-layer slabs of  Co hcp$(0001)$(a), Fe bcc$(001)$ and Ni fcc$(001)$ as a function of the Fermi level (in blue) from TB calculations. The corresponding number of electrons at the surface layer $N_e$ is shown in red. For convenience we have subtracted the number of valence electrons such that zero corresponds to the charge neutrality point. A zoom around the neutrality point is shown in the inset.} 
\end{figure}

In the same way that we have calculated the MCA as function of the band filling (or rather Fermi energy) in bulk Co hcp, one can look at the evolution of the surface component of the MCA when varying the electron filling of the surface plane. It has been shown that the application of an electric field perpendicular to the surface of a metal can significantly affect the surface MCA \cite{Duan2008} and therefore it is very relevant to study the variation of the surface MCA with the number of electrons in the surface layer. Indeed, the main effect of the electric field at the surface of a metal is the creation of a surface charge that can be simulated by changing the number of electrons in the outermost layer. In Fig. \ref{fig:MCA-filling} we show the outermost component of MCA as a function of the energy level together with the corresponding number of electrons for Co hcp$(0001)$, Fe bcc$(001)$ and Ni fcc$(001)$. As for bulk Co hcp we observe rapid variations of the MCA with the Fermi energy. Note that for Co and Fe at the neutrality point the slope of the MCA is negative while it is positive for Ni. This means that positively charging by a small amount the surface layer ({\sl i.e.} decreasing the number of electrons) will increase the MCA for Co and Fe and decrease for Ni. At this point it is important to have orders of magnitude in mind. Since the surface charge density (per surface unit) $\sigma$ is related to the perpendicular electric field $E$ by the relation $E=\frac{\sigma}{\epsilon_0}$ it comes out that an outward ({\sl i.e.} pointing out of the surface) electric field as large as 1V/{\AA} corresponds to a depletion of approximately 0.05 electron per surface atom. The corresponding variation of Fermi energy is approximately 0.04eV, 0.025 and 0.03eV for Co Fe and Ni respectively. In the insets of Fig. \ref{fig:MCA-filling} we have shown a zoom around the neutrality point over a realistic range of energy $\pm 0.1$eV which goes in line with the results of Duan {\sl et. al.}\cite{Duan2008} who showed by DFT that the effective application of an outward electric field increases the surface component of MCA for Fe and Co and decreases it for Ni\cite{comment}. Note, that this electric field effect could be amplified if instead of considering a surface we consider an interface with a material of high permittivity. 

\begin{figure}[!ht]
\includegraphics[width=6cm]{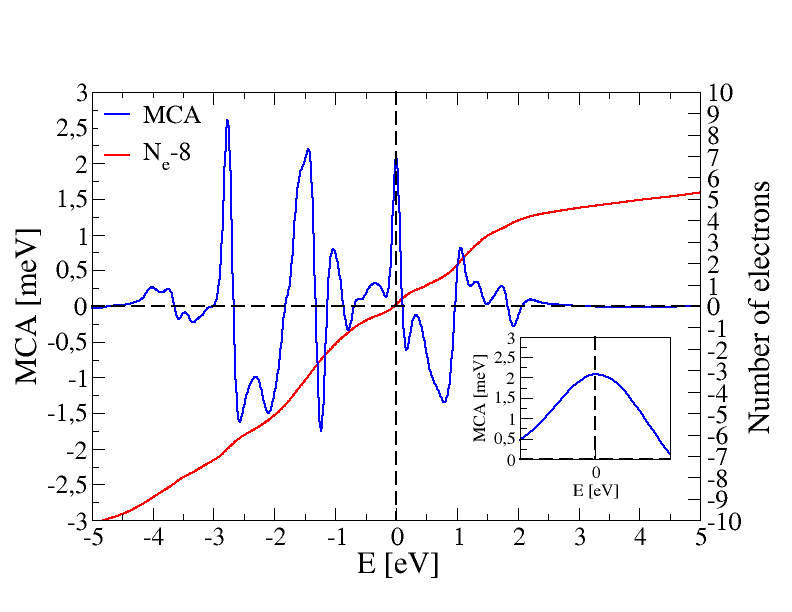}
	\caption{\label{fig:MCA-filling-4LFe}  total MCA of a 4-layer slab of Fe bcc$(001)$ as a function of the Fermi level (in blue) from TB calculations. The corresponding number of electrons (minus the number of valence electrons) is shown in red. A zoom around the neutrality point is shown in the inset.  Note that the value of the MCA at the maximum is lower than the one of Fig. \ref{fig:MCA-N}. This is due to the use of a Fermi-Dirac broadening to smooth the curve while a Marzari-Vanderbilt cold smearing was used to obtain the results of Fig. \ref{fig:MCA-N}}
\end{figure}

Finally it should be pointed out that in the case where the MCA is maximum (resp. minimum) at  the neutrality point, any electric field (outward or inward) will lead to a decrease (resp. increase) of the MCA. This is the case of the 4 layer slab of Fe bcc$(001)$ illustrated in Fig. \ref{fig:MCA-filling-4LFe} presenting a sharp maximum at the Fermi level.
This specific behaviour is induced by  finite size effect and is at the origin of the maximum of the MCA observed at 4 layers (Fig. \ref{fig:MCA-N}).

\section{Conclusion}
To summarize we have presented a comprehensive electronic structure analysis of the magneto-crystalline anisotropy energy of Fe bcc, Co fcc (and hcp) and Ni fcc slabs, obtained from three different codes: QE, QATK and TB. We have used the Force Theorem and its grand canonical formulation to define the layer resolved MCA. We show that total MCA is often strongly oscillating with the number of layers and this mainly originates from the contribution of inner (bulk-like) layers due to electronic confinement effects. However in most cases only the two or three outermost layers (depending on the surface packing density) are significantly perturbed. This allows to define the surface contribution to the MCA (Eq. \ref{eq:MCA-surf}). We also highlight an extremely rich and complex behaviour of the MCA in $k$-space and with the electron filling. Rapid variations are observed with change of sign and large amplitudes of the MCA.
 In particular we show that from the variation of the MCA surface component with electron filling one can predict the amplitude and the sign of the response to an applied electric field. Finally we hope that our work can provide an interesting benchmark and general trends for the design of materials with optimized magnetic properties. However on should keep in mind that any electronic structure determination of the MCA should be tested carefully with respect  to the computational ($k$ points, broadening etc..) and physical (Fermi level, lattice constant) parameters.

\acknowledgments
This project has received funding from the European Union’s Horizon 2020 research and innovation program under grant agreement No 766726.
\clearpage

\bibliographystyle{apsrev}

\end{document}